\documentclass[prd,aps,floats,preprintnumbers,preprint]{revtex4}

\usepackage{graphicx}
 
\newcommand{\be}{\begin{equation}}
\newcommand{\ee}{\end{equation}}
\newcommand{\bea}{\begin{eqnarray}}
\newcommand{\eea}{\end{eqnarray}}

\begin{document}


\setlength{\unitlength}{1mm}

\title{Inhomogeneous cosmological models and $H_0$ observations}

\author{Antonio Enea Romano$^{1,2,3,4}$}
\affiliation{
${}^1$Instituto de Fisica, Universidad de Antioquia, A.A.1226, Medellin, Colombia\\
${}^2$Department of Physics, National Taiwan University, Taipei 10617, Taiwan, R.O.C.\\
${}^3$Leung Center for Cosmology and Particle Astrophysics, National Taiwan University, Taipei 10617, Taiwan, R.O.C.\\
${}^4$Perimeter Institute, Waterloo, Canada\\
}



\begin{abstract}
We address some recent erroneous claim that $H_0$ observations are difficult to accommodate with LTB cosmological models, showing how to construct solutions in agreement with an arbitrary value of $H_0$ by re-writing the exact solution in terms of dimensionless parameters and functions. This approach can be applied to fully exploit LTB solutions in designing models alternative to dark energy without making any restrictive or implicit assumption about the inhomogeneity profile. 
The same solution can also be used to study structure formation in the regime in which perturbation theory is not enough and an exact solution of the Einstein's equation is required, or to estimate the effects of a local inhomogeneities on the apparent equation of state of dark energy.
\end{abstract}

\maketitle

\section{Introduction}

Inhomogeneous solutions of the Einstein's field equations have received same attention as alternatives cosmological model to dark energy.
The main idea is that the space inhomogeneity is able to mimic in red-shift space the effects of homogeneous dark energy, without the need of modifying gravity or introducing a model for dark energy.
Some interesting theoretical investigation about LTB  or other inhomogeneous solutions in a cosmological context can be found for example in
\cite{Romano:2006yc,Chung:2006xh,Yoo:2008su,Alexander:2007xx,Alnes:2005rw,GarciaBellido:2008nz,GarciaBellido:2008gd,GarciaBellido:2008yq,February:2009pv,Quercellini:2009ni,Clarkson:2007bc,Ishibashi:2005sj,Bolejko:2011ys,Clifton:2009kx,Zibin:2011ma,Bull:2011wi,Romano:2009mr,Romano:2007zz,Romano:2009ej,Romano:2009qx,Valkenburg:2011tm,Bolejko:2012uj,Bolejko:2012ue,Hoyle:2012pb,Mustapha:1998jb,Romano:2009xw,Celerier:1999hp,Hellaby:2009vz,Dabrowski:1995ae,Dabrowski:2004hx,Dabrowski:1986,jdb1,jdb2,jdb3}.

Another interesting application of LTB solutions is to consider them as realistic model of the non-linear effects of large scale inhomogeneities in presence of dark energy \cite{Romano:2010nc,Sinclair:2010sb,Romano:2011mx,Romano:2012gk,Marra:2012pj,Marra:2010pg,Valkenburg:2011ty,Valkenburg:2012ds,arXiv:1106.1611,Clifton:2008hv}.
There have been several attempts to fit cosmological data  \cite{Biswas:2010xm,Moss:2010jx}, and recently \cite{Riess:2011yx} it has been claimed that accurate measurements of $H_0$ are sufficient to rule out best fit void models. 
In this paper we will show how such claims are erroneous, and it is easy to construct general LTB models which can accommodate any observed value of $H_0$ value, arguing that the large residual functional freedom should be fully exploited before making conclusive statements about their compatibility with observations.
Our approach differ from previous ones because $H_0$ is not a quantity derived from other parameters of the model, but rather the fundamental natural physical scale in terms of which all other dimensionful quantities are expressed.
The paper is organized as following.
Using a generalized conformal time variable we derive in a unified and compact way the solution with and without cosmological constant , showing how the latter one reduces to the first one in the appropriate limit.
We then introduce dimensionless functions defining the LTB solution, expressing the analytical solution in a form in which the observed $H_0$ is a manifestly free length scale parameter.
The analytical solutions we derive could have different applications; they could for example be  be used to fit observational cosmological data or for an exact study of the spherical collapse in structure formation.

\section{Exact solution for LTB models}
The LTB solution can be written
as \cite{Lemaitre:1933qe,Tolman:1934za,Bondi:1947av} as
\begin{eqnarray}
\label{LTBmetric} %
ds^2 = -dt^2  + \frac{\left(R,_{r}\right)^2 dr^2}{1 + 2\,E(r)}+R^2
d\Omega^2 \, ,
\end{eqnarray}
where $R$ is a function of the time coordinate $t$ and the radial
coordinate $r$, $E(r)$ is an arbitrary function of $r$, and
$R_{,r}=\partial_rR(t,r)$.

The Einstein equations with dust and a cosmological constant give
\begin{eqnarray}
\label{eq2} \left({\frac{\dot{R}}{R}}\right)^2&=&\frac{2
E(r)}{R^2}+\frac{2M(r)}{R^3}+\frac{\Lambda}{3} \, , \\
\label{eq3} \rho(t,r)&=&\frac{2M,_{r}}{R^2 R,_{r}} \, ,
\end{eqnarray}
with $M(r)$ being an arbitrary function of $r$, $\dot
R=\partial_tR(t,r)$ and $c=8\pi G=1$ is assumed throughout the
paper. Since Eq. (\ref{eq2}) contains partial derivatives respect
to time only, its general solution can be obtained from the FLRW
equivalent solution by making every constant in the latter one an
arbitrary function of $r$.

An analytical solution can be found by introducing
a new coordinate $\eta=\eta(t,r)$ and a variable $a$ by \bea
\left(\frac{\partial\eta}{\partial
t}\right)_r=\frac{r}{R}\equiv\frac{1}{a}\,, \label{etadef} \eea
and new functions by 
\bea
\rho_0(r)\equiv\frac{6 M(r)}{r^3}\,,\quad
k(r)\equiv-\frac{2E(r)}{r^2}\,. \eea 
Then Eq. (\ref{eq2}) becomes
\be \left(\frac{\partial a}{\partial\eta}\right)^2 =-k(r)
a^2+\frac{\rho_0(r)}{3} a+\frac{\Lambda}{3} a^4\,, \label{FriedGen} 
\ee 
where $a$ is now
regarded as a function of $\eta$ and $r$, $a=a(\eta,r)$. It should
be noted that the coordinate $\eta$, which is a generalization of
the conformal time in a homogeneous FLRW universe, has been only
implicitly defined by Eq.~(\ref{etadef}). The actual relation
between $t$ and $\eta$ can be obtained by integration once $a(\eta,r)$ is known:
\be
t(\eta,r)=\int_0^{\eta}{a(x,r)dx}+t_b(r) \,, \label{teta}
\ee
which can be computed analytically, and involve elliptic integrals of the third kind\cite{ellint}.

The function $t_B(r)$ plays the role of constant of integration, and is an arbitrary function of $r$, sometime called bang function, since by construction at time $t=t_b(r)$ we have $a(t_b(r),r)=0$, and correspond to the fact that the big bang initial singularity can happen at different times at different positions from the center in a LTB space. 

\subsection{Case with no cosmological constant}

When $\Lambda=0$ the eq.(\ref{FriedGen}) can be easily solved, and the solution of physical interest satisfying the the big bang initial condition $a(0,r)=0$ is :

\bea
\label{LTB soln2 R} a(\eta,r) &=& \frac{\rho_0(r)}{6k(r)}
     \left[ 1 - \cos \left( \sqrt{k(r)} \, \eta \right) \right]=\frac{\rho_0(r)\sin^2 \left(\frac{1}{2}\sqrt{k(r)} \, \eta \right)}{3k(r)} \, ,\label{sol}
\eea
and after integration, using eq.(\ref{teta})
\bea     
\label{LTB soln2 t} t(\eta,r) &=& \frac{\rho_0(r)}{6k(r)}
     \left[ \eta -\frac{1}{\sqrt{k(r)}} \sin
     \left(\sqrt{k(r)} \, \eta \right) \right] + t_{b}(r) \, .
\eea

It should be noted that this solution is valid for any sign of the the function $k(r)$, and is more convenient than the equivalent two branches form used by some other authors \cite{Celerier:2009sv}.

\subsection{Case with cosmological constant} 

The general analytical solution for a FLRW model with dust and cosmological
constant was obtained by Edwards \cite{Dilwyn} in terms of elliptic functions.
By an appropriate choice of variables and coordinates, we may extend it
to the LTB case thanks to the spherical symmetry of both LTB and FLRW models,
and to the fact that dust follows geodesics without being affected by
adjacent regions.
Inspired by the construction of the solution for the FLRW case and by dimensional analysis we can make the ansatz
\bea
\tilde{a}(T,r)&=&\frac{\beta }{\delta  K(r)+\gamma\tilde {\phi} (T \alpha )}\,,
\eea
where we are using the following dimensionless quantities:
\bea
k(r)&=&(a_0 H_0)^2 K(r)\,, \\
\eta&=&T(a_0 H_0)^{-1}\,,\\
\rho_0(r)&=&3 \Omega^0_M(r) a_0^3 H_0^2\,,\\
\Lambda&=&3 \Omega_{\Lambda}H_0^2\,, \\
a(\eta,r)&=&a(T(a_0 H_0)^{-1},r)=\tilde{a}(T,r) \,, \\
\frac{\partial a(\eta,r)}{\partial \eta}&=&a_0 H_0 \frac{\partial \tilde{a}(T,r)}{\partial T}=a_0 H_0\tilde{a}'(T,r)\,,
\eea 
and we are denoting with a tilde the function $\tilde{\phi}$ for consistency with the notation for the scale factor, i.e. we use a tilde for quantities
which depend on the dimensionless generalized conformal time variable $T$.
After substituting the above ansatz in the Einstein's equation we get a differential equation equation of the form:
\be
A\tilde{\phi}'(T)^2+B\tilde{\phi}(T)^3+C\tilde{\phi}(T)^2+D\tilde{\phi}(T)+E=0 \,, \label{EW}
\ee 
where  the coefficients $\{A,B,C,D\}$ depend on the parameters $\{\alpha,\beta,\gamma,\delta\}$.
 
If we now impose the conditions
\bea
\frac{B}{A}&=&-4 \,\\
C&=&0 \,\\
D\over A&=&g_2 \, \\
E\over A&=&g_3 \,
\eea
the Einstein's equation (\ref{EW}) reduces to the canonical form of the Weierstrass differential equation 
\be
\left(\frac{d\tilde{\phi}}{dT}\right)^2=4\tilde{\phi}^3-g_2\tilde{\phi}-g_3\, ,\label{eqweir} 
\ee 
whose solution is an elliptic function conventionally expressed using the notation $\tilde{\phi}(T,g_2,g_2)$.
The above conditions give :
\bea
\beta&=&\frac{a_0\Omega^0_M(r)\gamma }{4 \alpha ^2}\,, \\
\delta&=&\frac{\gamma }{12 \alpha ^2} \,,\\
g_2&=&\frac{K(r)^2}{12 \alpha ^4} \,,\\
g_3&=&\frac{2 K(r)^3-27 \Omega_{\Lambda} (\Omega^0_M(r))^2}{432 \alpha ^6} \,.
\eea
Since we have only two conditions for the four unknowns $\{\alpha,\beta,\gamma,\delta\}$ the ansatz we made allows to express the solution of the Einstein's equation in terms of the Weierstrass elliptic functions for an arbitrary choice of two of them.
We choose :
\be
\{\alpha=1\,;\,\beta=3 a_0\Omega^0_M(r)\,;\,\gamma=12\,;\,\delta= 1\} \,, \label{solL}
\ee
which gives the solution
\bea
\tilde{a}(T,r)&=&\frac{3 a_0\Omega^0_M(r)}{K(r)+12 \tilde{\phi} (T,g_2(r),g_3(r))} \,,\\
g_2(r)&=&\frac{K(r)^2}{12} \,,\\
g_3(r)&=& \frac{1}{432} \left(2 K(r)^3-27 \Omega_{\Lambda} (\Omega^0_M(r))^2\right)\,,
\eea
or after multiplying every term by $(a_0 H_0)^2$ and using the original dimensionful quantities $\eta,k(r),\rho_0(r)$

\bea
a(\eta,r)&=&\frac{\rho_0(r)}{k(r)+12 \phi (\eta,g_2(r),g_3(r))}=\tilde{a}(T,r) \,,\\
\phi(\eta,r)&=&\tilde{\phi}(\eta (a_0 H_0),r)(a_0 H_0)^2=\tilde{\phi}(T,r)(a_0 H_0)^2\,. 
\eea

\subsection{Relation between the two solutions}
In the previous subsections we have derived the analytical solution of the Einstein's equation for a LTB model with and without cosmological constant.
Since we are using the same coordinate system to derive the two solutions we expect that eq.(\ref{solL}) should reduce to eq.(\ref{sol}) when $\Lambda=0$.
In fact the discriminant of the Weierstrass function which solves eq.(\ref{FriedGen}) 
\bea
\Delta=g_2^3-27 g_3^2=\frac{1}{256} \Omega_{\Lambda} (\Omega^0_M(r))^2 \left(4 K(r)^3-27\,, \Omega_{\Lambda} (\Omega^0_M(r))^2\right)\,,
\eea
is zero when the cosmological constant vanishes.

It can be shown that when the discriminant is zero the Weierstrass elliptic function can be expressed in terms of trigonometric functions :
\bea
\phi(T,3 g_3^{2/3},g_3)&=&\frac{1}{2}g_3^{1/3}\left[3 \csc ^2\left(\frac{1}{2}T\sqrt{6 g_3^{1/3}} \right)-1\right]\,.
\eea
If we now substitute the above equation in eq.(\ref{solL}), after some manipulation we get exactly eq.(\ref{LTB soln2 R}), which can be considered a consistency check for the solutions derived so far.
In the rest of the paper we choose the so called FLRW gauge, i.e. the coordinate system in which $\rho_0(r)$ is constant. 

\section{LTB solutions and $H_0$ observations}
The exact solutions derived above are not defined in terms of directly observables quantities, since we have not introduced explicitly any system of units.
If we are interested in cosmological applications the natural scale to choose is $H_0^{-1}$, so it is convenient to re-write equation (\ref{FriedGen}) in the familiar FLRW-like form:
\bea
H^2(t,r)&=&H_0^2\Big[-K(r)\Big(\frac{a_0}{a}\Big)^2+\Omega^0_M(r)\Big(\frac{a_0}{a}\Big)^3+\Omega_{\Lambda}\Big] \,, \\
&=&H_0^2\Big[\Omega_K(t,r)+\Omega_M(t,r)+\Omega_{\Lambda}\Big] \,, 
\eea
where we have define 
\bea
H(t,r)&=&\frac{\dot{a}}{a} \,,\\
H_0&=&H(t_0,0) \,,\\
k(r)&=&(a_0 H_0)^2 K(r)\,, \\
K(r)&=&-\Omega_K(r) \,, \\
\Omega_K(t,r)&=&-K(r)\Big(\frac{a_0}{a}\Big)^2 \,, \\
\Omega_M(t,r)&=&\Omega^0_M(r)\Big(\frac{a_0}{a}\Big)^3 \,,\\
\rho_0(r)&=&3 \Omega^0_M(r) a_0^3 H_0^2\,,\\
\Lambda&=&3 \Omega_{\Lambda}H_0^2\,.
\eea
In this form the observed $H_0$ appears explicitly as one of the parameters defining the LTB model, and all the other quantities are dimensionless.

The exact solution can then be conveniently re-written in terms of dimensionless functions and coordinates as
\bea
a(T,r)&=&\frac{a_0 \Omega^0_M(r) \sin ^2\left(\frac{1}{2} T \sqrt{K(r)}\right)}{K(r)} \,,\\
t(T,r) &=& H_0^{-1}\frac{\Omega^0_M(r)}{2K(r)}
     \left[ T -\frac{1}{\sqrt{K(r)}} \sin
     \left(\sqrt{K(r)} \, T \right) \right] + t_{b}(r) \,,\\
\eta&=&T(a_0 H_0)^{-1}\,.\\
\eea
After defining the Hubble rate in terms of the generalized conformal time variable $\eta$ as 
\bea
H^{LTB}&=&\frac{\partial_t a(t,r)}{a(t,r)}=\frac{\partial_{\eta} a(\eta,r)}{a(\eta,r)^2}= (a_0 H_0)\frac{\tilde{a}'(T,r)}{a(T,r)^2}\,. \label{HLTB}\\
\eea
if we want the exact solution to be in agreement with $H_0$ observations we can impose the two following conditions 
\bea
a(\eta_0,0)&=&a_0 \,, \label{a0}\\
H^{LTB}(\eta_0,0)&=&H_0 \label{H0}\label{H0}\,,
\eea
where $a_0$ is, as expected, an arbitrary parameter, and $\eta_0$ is the value of the generalized conformal time coordinate $\eta$ corresponding to the central observer today, i.e. satisfying 
\bea
t(\eta_0,0)=t_0 \,.
\eea
We will discuss the consequences of the above conditions in the two following sections according to the presence or absence of the cosmological constant term in the Einstein's equations.
As mentioned at the end other previous section, we will will use the freedom in the choice of the radial coordinate to fix the coordinates in which $\rho_0=const.$, and consequently $\Omega^0_M(r)=const.$
\section{Vanishing cosmological constant case}
In the vanishing cosmological constant case we get
\bea
\frac{a_0 \Omega^0_M f_0^2}{K_0 (f_0^2+1)}=a_0\,,\\
\frac{H_0 K_0^{3/2} \left(f_0^2+1\right)}{\Omega^0_M f_0^3}=H_0\,, \label{H0}
\eea
where 
\bea
K_0&=&K(0) \,, \\
T_0&=&\eta_0(a_0 H_0)^{-1}\,,\\
f_0&=&\tan{\left(\frac{\sqrt{K_0}T_0}{2}\right)} \label{f0}.
\eea
and we have kept $a_0,H_0$ on both side of the equations just to show their relation with eqs.(\ref{a0},\ref{H0}), but they obviously cancel.
The above equations can be solved for $f_0,\Omega^0_M$ to get
\bea
\Omega^0_M&=&1+K_0\,,\\
f_0&=&\sqrt{K_0}\,.
\eea
or equivalently, after inverting eq.(\ref{f0})
\bea
\Omega^0_M&=&1+K_0\,,\\
T_0&=&\frac{2\arctan{(f_0)}}{\sqrt{K_0}}=\frac{2\arctan{(\sqrt{K_0})}}{\sqrt{K_0}}\,.
\eea
The meaning of the above relations is that for an arbitrary value of $K_0$ there is always a LTB solution in agreement with $H_0$ observations given by
\bea
a(T,r)&=&\frac{a_0 (K_0+1) \sin ^2\left(\frac{1}{2} T \sqrt{K(r)}\right)}{K(r)} \,,\\
t(T,r) &=& H_0^{-1}\frac{1+K_0}{2K(r)}
     \left[ T -\frac{1}{\sqrt{K(r)}} \sin
     \left(\sqrt{K(r)} \, T \right) \right] + t_{b}(r) \,,\\
T_0&=&\frac{2\arctan{(\sqrt{K_0})}}{\sqrt{K_0}}\,.
\eea
As expected $a_0$ does not appear in observable quantities such as the cosmic time $t(\eta,r)$.
This solution will be by construction in agreement with an arbitrary value of $H_0$, while the residual functional freedom in $K(r)$ could be exploited to fit other observables such as the luminosity distance. We will report about this in a separate work, in particular in relation to previous failed attempts to fit both the luminosity distance and the cosmic microwave radiation (CMB). What we have found is in agreement with a previous investigation based on a local Taylor expansion \cite{arXiv:1105.1864}, and provides a useful method to design LTB models without cosmological constant to fit observational data.
\section{Not vanishing cosmological constant case}
 The case in which the cosmological constant is not vanishing can be interesting because it provides a simple way to model spherically symmetric large scale inhomogeneities around a central observer, which, as discussed in  \cite{arXiv:1104.0730} are expected to be the dominant ones due to other strong evidences of isotropy such as the CMB radiation.
After re-expressing the metric using the same dimensionless quantity introduced above we get:
\bea
\frac{3 a_0^3 H_0^2 {\Omega_M}^0}{a_0^2 H_0^2 K_0+3 \tilde{\phi}_0 } &=& a_0\,, \\
-\frac{4 H_0 \tilde{\phi}'_0}{{\Omega_M}^0}&=&H_0\,,
\eea
where 
\bea
\tilde{\phi}_0&=&\tilde{\phi}(T_0,0)\,,\\ \label{phi0}
\tilde{\phi}'_0&=&\left(\frac{\partial \tilde{\phi}}{\partial T}\right)_{\{{T=T_0,r=0}\}}\,,\\
K_0&=& K(0).
\eea
Our final goal is to determine $\Omega_M,T_0$ from the above equation, but apparently we have an extra unknown, i.e. 
$\tilde{\phi}'_0$. This is not really independent from $\phi_0$, since the Weierstrass equation allows to express algebraically one in terms of the other.
This is a general property which can be used whenever we have to  simplify or manipulate analytical expressions involving the solution we have derived in terms of elliptic functions:

\bea
\tilde{\phi}'_0&=&\sqrt{-\frac{K_0^3}{216}-\frac{K_0^2 \phi_0}{12}+\frac{\Omega_{\Lambda} (\Omega_M^0)^2}{16}+4 \tilde{\phi}_0^3}.
\eea

Finally the solution of the system of equations.(\ref{a0},\ref{H0}) gives:

\bea
{\Omega_M^0}&=&K_0-\Omega_{\Lambda}+1\,, \\
\tilde{\phi}_ 0&=& \frac{1}{12} (2 K_0-3 \Omega_{\Lambda}+3)\,,
\eea
Another possible way to interpret the solutions we have derived is to consider the Einstein's equation at the center which gives
\be
1=\Omega^0_M(0)+\Omega_K+\Omega_{\Lambda} \,,
\ee
where we are using the notation
\be
\Omega_K=-K(0)\,,
\ee
since locally at the center the $LTB$ solution is equivalent of a $FLRW$ solution with the above value of $\Omega_K$.
This is in fact equivalent to the equation(\ref{H0}), since the solution of the Einstein's equation substituted into the definition of $H^{LTB}$ in eq.(\ref{HLTB}) gives the Einstein's equation itself. According to this interpretation the Einstein's equation at the center gives the 
condition to determine $\Omega^0_M$ in agreement with the $H_0$ observations.
Using the equations derived above, for a given value of $\Omega_{\Lambda},K_0$ we can now determine $\Omega^0_M$ and $\tilde{\phi}_0$ which will give a model in agreement with an arbitrary observed $H_0$. We can finally solve eq.(\ref{phi0}) to get $T_0$ or $\eta_0$. While in the vanishing cosmological constant case this can be done analytically, in this case we need to solve numerically the equation to get $T_0$.  

\section{Possible applications}
One of the main problems that LTB models without dark energy seem to face in fitting observational data is the value of $H_0$ \cite{Biswas:2010xm,Moss:2010jx}.
The approach normally adopted previously consisted in defining a void model in terms of a limited set of parameters and then indirectly derive the value of $H_0$. In other words $H_0$ was not considered as a fundamental scale in terms of which to determine the geometry of the LTB solution as we are proposing here. In our approach instead $H_0$ is the primary ingredient in defining the LTB model, and this should help to improve the fitting of these model to observations. In order to fit CMB data a LTB model should have the same distance to the last scattering surface of the best fit $\Lambda CDM$ model, and have the same value of $H(z_{LS})$ at the last scattering surface.
It is also known from a numerical inversion method \cite{Celerier:2009sv}, that it is possible to construct a LTB model able to exactly mimic both the luminosity distance $D_L(z)$ and the $H(z)$ of a $\Lambda CDM$ model, but this requires the use of both $k(r)$ and $t_b(r)$. This implies that allowing for a inhomogeneous big bang we can expect to fit both luminosity distance and the CMB, and in fact previous claims of the impossibility to fit $CMB, H_0$ and $D_L(z)$ were based on void models with $t_b(r)=0$. 
Since using the solution in the form we have derived gives a LTB model in agreement with any value of $H_0$, the problem will probably be in getting the value of $H(z_{LS})$ in agreement with CMB observations, but a conclusive answer to this question can be given only by fitting the data with models which have more freedom in defining the inhomogeneity profile, while the void models analyzed so far e only depend on few parameters. In this regard a differential inversion method would be more appropriate since it would allow to fully exploit the functional degree of freedom of LTB solutions. his application is left to a separate work.
The important point is that contrary to previous claims to rule out void models using $H_0$ observations, the correct way to design a LTB model is to introduce $H_0$ from the beginning and express the other parameters and functions such as $K(r)$ in terms of it, rather than deducing it.
The indirect method used so far in fitting cosmological data can in fact artificially limit the genuine freedom of LTB solutions according to the particular ansatz made for the inhomogeneity profile.    
\section{Conclusions}
We have shown how to construct a LTB solutions in agreement with any value of $H_0$ by using it as the fundamental scale in terms of which to define all the other quantities determining the model, proving that any claims that $H_0$ measurements are enough to rule out LTB models are wrong.
This will allow to fit data in a more efficient way, since $H_0$ does not need to be computed from the other parameters of the model, but on the contrary can be fixed directly without any additional computational effort. In this way it will be possible to fully explore the space of LTB solutions as viable cosmological model, going beyond the limited class of models examined so far.

Another natural application is in relation to structure formation, where the derived solutions could be used to study exactly the spherical collapse in presence of a cosmological constant.
In this context it could actually be more convenient to define the solution directly in terms of the density contrast, and we leave this to a future work.
The exact solution could also be used in numerical simulations of structure formation in the regime in which the collapse of compact objects cannot be treated accurately with perturbation theory, and an exact solution of the Einstein's equations is required.

\begin{acknowledgments}
We thank A. Starobinsky, M. Sasaki for useful comments and discussions.
A. E. Romano was supported by CODI project IN615CE and the "Dedicacion Exclusiva" program of the University of Antioquia.
\end{acknowledgments}

\end{document}